\documentclass{aa}  

\usepackage{graphicx}
\usepackage{txfonts}
\usepackage{xcolor}
\usepackage[colorlinks=true,citecolor=blue]{hyperref}
\usepackage{comment}
\newcommand{\ixpe}{\text{IXPE}}

\newcommand{\rxte}{\text{RXTE}}

\newcommand{\nustar}{\text{NuSTAR}}

\newcommand{\nicer}{\text{NICER}}
\newcommand{\maxi}{\text{MAXI}}

\newcommand{\fluxcgs}{erg~s$^{-1}$~cm$^{-2}$}
\newcommand{\lumcgs}{erg~s$^{-1}$}
\newcommand{\percm}{\,cm$^{-2}$}        
\newcommand{\pers}{\,s$^{-1}$}  

\newcommand{\xspec}{\textsc{xspec}}
\newcommand{\diskbb}{\texttt{diskbb}}
\newcommand{\bbodyrad}{\texttt{bbodyrad}}

\newcommand{\relxillns}{\texttt{relxillns}}
\newcommand{\polconst}{\texttt{polconst}}
\newcommand{\tbabs}{\texttt{tbabs}}

\newcommand{\thcomp}{\texttt{thcomp}}

\begin{document} 

    \title{X-ray spectropolarimetry of the bright atoll Serpens~X-1}
    \author{F. Ursini\inst{1},
            A. Gnarini\inst{1},
            S. Bianchi\inst{1},
            A. Bobrikova\inst{2},
            F. Capitanio\inst{3},
            M. Cocchi\inst{4},
            S. Fabiani\inst{3},
            R. Farinelli\inst{5},
            P. Kaaret\inst{6},
            G.~Matt\inst{1},
            M. Ng\inst{7},
            J. Poutanen\inst{2},
            S. Ravi\inst{7},
          \and
          A. Tarana\inst{3}
          }

    \institute{Dipartimento di Matematica e Fisica, Università degli Studi Roma Tre, via della Vasca Navale 84, 00146 Roma, Italy\\ \email{francesco.ursini@uniroma3.it}
    \and
    Department of Physics and Astronomy, FI-20014 University of Turku, Finland
    \and
    INAF Istituto di Astrofisica e Planetologia Spaziali, Via del Fosso del Cavaliere 100, 00133 Roma, Italy
    \and
    INAF Osservatorio Astronomico di Cagliari, via della Scienza 5, I-09047 Selargius (CA), Italy
    \and
    INAF Osservatorio di Astrofisica e Scienza dello Spazio di Bologna, Via P. Gobetti 101, I-40129 Bologna, Italy
    \and
    NASA Marshall Space Flight Center, Huntsville, AL 35812, USA
    \and
    MIT Kavli Institute for Astrophysics and Space Research, Massachusetts Institute of Technology, Cambridge, MA 02139, USA}

\titlerunning{X-ray polarization of Ser X-1} 
\authorrunning{F. Ursini et al.} 

   \date{Received ...; accepted ...}

 
\abstract{
We present simultaneous X-ray polarimetric and spectral observations of the bright atoll source Ser~X-1 carried out with the Imaging X-ray Polarimetry Explorer (\ixpe), \nicer,\, and \textit{\nustar}. We obtain an upper limit of 2\% (99\% confidence level) on the polarization degree in the 2--8 keV energy band. We detect four type-I X-ray bursts, two of which during the \ixpe\ observation. This is the first time that has IXPE observed type-I X-ray bursts, and it allows us to place an upper limit on their polarization degree; however, due to the limited total number of counts in each burst, we obtain a relatively high upper limit (80\%). We confirm the presence of reflection features in the X-ray spectrum, notably a broad iron line. Fitting the data with a relativistic reflection model, we derive a disk inclination of 25\degr. The spectral and polarization properties are comparable with other atolls observed by \ixpe, suggesting a similar accretion geometry, and the relatively low polarization is consistent with the low inclination.  
  }


   \maketitle
%
\section{Introduction}
Accreting, weakly magnetized neutron star low-mass X-ray binaries (NS-LMXBs) are very bright X-ray sources and excellent laboratories for studying the physics of accretion onto compact objects. They are divided into two main classes, Z and atoll, according to the pattern they trace on X-ray color-color diagrams \citep{hasinger&vanderklis}. Atoll sources are also less luminous and tend to have weaker magnetic fields than Z sources. The X-ray spectrum of NS-LMXBs is generally well described by a soft thermal component plus a hard Comptonization component, but their physical origin is still uncertain. The soft component could be due to either the accretion disk or the NS, while the hard component could originate in a hot corona, in a boundary layer (BL) between the disk and the NS \citep{popham&sunyaev2001}, or in a spreading layer (SL) around the NS \citep{inogamov&sunyaev1999}.

Thanks to X-ray polarimetric results from the Imaging X-ray Polarimetry Explorer (\ixpe; \citealt{weisskopf2022,soffitta2021}), our understanding of NS-LMXBs is advancing significantly. So far, \ixpe\ has detected significant polarization in at least eight NS-LMXBs, both atoll and Z sources \citep{farinelli2023,cocchi2023,jayasurya2023,chatterjee2023,ursini2023,dimarco2023_4U1820,fabiani2024,saade2024,lamonaca2024,bobrikova2024}, plus one stringent upper limit \citep{capitanio2023}. Detailed spectropolarimetric analysis has revealed that the hard component dominates the polarization signal, while the soft emission has a low polarization \citep[][]{farinelli2023,cocchi2023,ursini2023,dimarco2023_4U1820,lamonaca2024}. The results so far suggest that the soft thermal component is due to the accretion disk, while Comptonization likely takes place in a BL or SL. In principle, reflection off the disk surface can significantly contribute to the observed polarization \citep{lapidus&sunyaev1985}. 
However, it is still unclear whether 
the reflected component is the dominant contributor \cite[][]{ursini2023,fabiani2024,saade2024}. For example, in the case of the atoll source GX~9+9, the observed polarization is likely a combination of Comptonization and reflection \citep{ursini2023}. Interestingly, \ixpe\ measured surprisingly large polarization in other atolls in the soft state, up to $6\% \pm 2\%$ in the 6--8 keV band in 4U~1624$-$49 \citep{saade2024} and $10\% \pm 2\%$ in the 7--8 keV band in 4U~1820$-$30 \citep{dimarco2023_4U1820}. These values are much higher than expected for any plausible BL or SL geometry \citep{gnarini2022,capitanio2023,farinelli2024}. 
A possible explanation for the high polarization is a slab-like accretion disk corona or a strong contribution due to scattering off the disk or a wind \citep{saade2024}. 

Serpens X-1 (Ser~X-1) is a bright, persistent atoll NS-LMXB. It is a well-studied source, consistently observed in the high luminosity, soft spectral (``banana'') state \citep[e.g.,][]{oosterbroek2001,masetti2004,cackett2008,chiang2016_a,matranga2017,ludlam2018_SerX1,mondal2020_SerX1}. 
The occurrence of type-I X-ray bursts is well established in this source \citep{sztajno1983,balucinska1985,galloway2008}. From observations with the Rossi X-ray Timing Explorer (\rxte), \cite{galloway2008} found evidence of photospheric radius-expansion bursts, and from their peak luminosity they estimated a distance of $7.7 \pm 0.9$ kpc.
The X-ray spectrum shows both a relativistically broadened Fe K$\alpha$ emission line \citep{bhattacharyya2007,cackett2008,cackett2010,miller2013,chiang2016_a,chiang2016_b} and a Compton reflection hump at 10--20 keV \citep{miller2013,matranga2017,ludlam2018_SerX1,mondal2020_SerX1}. Optical spectroscopy indicates that the Ser~X-1 system has a low binary inclination, less than 10\degr\  \citep{cornelisse2013}. Different values of the disk inclination, between $\sim$10\degr\ and $\sim$50\degr, have been reported based on X-ray reflection spectroscopy \citep{bhattacharyya2007,miller2013,matranga2017,ludlam2018_SerX1,mondal2020_SerX1}. In any case, Ser~X-1 is one of the X-ray-brightest atolls known \citep{asai2022} and thus an excellent target for X-ray polarimetry. In this paper we report on the first \ixpe\ observation of this source, performed jointly with the Neutron Star Interior Composition Explorer (\nicer) and \textit{\nustar}.

The paper is structured as follows. In Sect.~\ref{sec:data} we describe the observations and data reduction. In Sect.~\ref{sec:analysis} we present the X-ray spectropolarimetric analysis. In Sect.~\ref{sec:conclusions} we discuss the results and the main conclusions.
\section{Observations and data reduction}\label{sec:data}

\begin{table}
\caption{\label{tab:obs-log}
Log of the \ixpe\ and \nicer\ observations.
}
\begin{center}
  \begin{tabular}{cccc}
    \hline    \hline
 Satellite & Obs. Id. & Start time & Net exp. \\
 &&(UTC)&(ks)\\
\hline 
\ixpe & 03003901 & 2024-04-15T00:03:51 & 37.3 \\
\nicer & 7700020102 & 2024-04-14T18:36:45 & 7.8 \\
\textit{\nustar} & 91001316002 & 2024-04-15T08:51:09 & 33.8 \\
\hline 
\end{tabular}
\end{center}
\end{table}

\ixpe\ observed Ser~X-1 on 2024 April 14 with its three detector units (DUs)/mirror module assemblies (MMAs), for a net exposure time of 37.3 ks (Table \ref{tab:obs-log}). We produced cleaned level~2 event files using standard filtering criteria with the dedicated {\sc ftools} (v6.33) tasks\footnote{\url{https://heasarc.gsfc.nasa.gov/docs/ixpe/analysis/IXPE-SOC-DOC-009D_UG-Software.pdf}} and the latest calibration files (CALDB 20240125) and response matrices (v13). 
We extracted the Stokes $I$, $Q,$ and $U$ spectra from circular regions with a radius of 120\arcsec. We did not subtract the background, following the prescription by \cite{dimarco2023_bkg} for bright sources. However, we verified that background subtraction does not significantly alter the results;  
in particular, the polarimetric measurements are not affected at all. 
We performed the data analysis following the weighted scheme \textsc{neff} \citep{baldini2022,dimarco2022}.
Given the source brightness, each energy bin of the flux ($I$) spectra contains more than 40 counts, ensuring the applicability of the $\chi^2$ statistics. Thus, we did not re-bin the $I$ spectra but did apply a constant energy binning of 0.2 keV for $Q$ and $U$ Stokes spectra. We fitted the $I$, $Q$, and $U$ Stokes spectra from the three DU/MMAs independently. 

\nicer\ performed two observations of the source with continuous exposure in the period 2024 April 14-15 simultaneously with \ixpe, for a net exposure time of 7.8~ks. The \nicer\ data were reduced using \textsc{ftools} tasks and the processing script \textsc{nicerl2} to apply standard screening and calibration (CALDB 20240206). 
We used the data in the 1.5--10 keV band.

\textit{\nustar}\ \citep{harrison2013} observed the source with its X-ray telescopes on Focal Plane Module A (FPMA) and B (FPMB), with a net exposure of 33.8 ks and an overlap with the last $\sim$4~hr of the \ixpe\ exposure. We produced cleaned event files with the standard \texttt{nupipeline} task and the latest calibration files (CALDB 20240325). The \textit{\nustar}\ background is not negligible at all energies, and thus we performed background subtraction as follows. For both detectors, the background was extracted from a circular region with a standard radius of 60\arcsec. The source radius was set to 120\arcsec\ following a procedure that maximizes the signal-to-noise ratio \citep{pico2004}. 
Finally, we re-binned the spectra with the standard task \texttt{ftgrouppha} with the optimal scheme by \cite{kaastra&bleeker2016}, also requiring a minimum signal-to-noise of 3 in each bin. The FPMA and FPMB spectra were fitted independently and not co-added. We used the data in the 3--30~keV range since the background starts dominating above 30~keV. 

\section{Data analysis}\label{sec:analysis}
\subsection{Timing properties}
The X-ray flux of Ser~X-1 displays a quasiperiodic modulation on a timescale of decades \citep{asai2022}.\ However, the recent light curve of the {Monitor of All-sky X-ray Image} \cite[\maxi;][]{matsuoka2009} shows only a limited amplitude variability. 

\begin{figure}
\includegraphics[width=1.0\columnwidth]{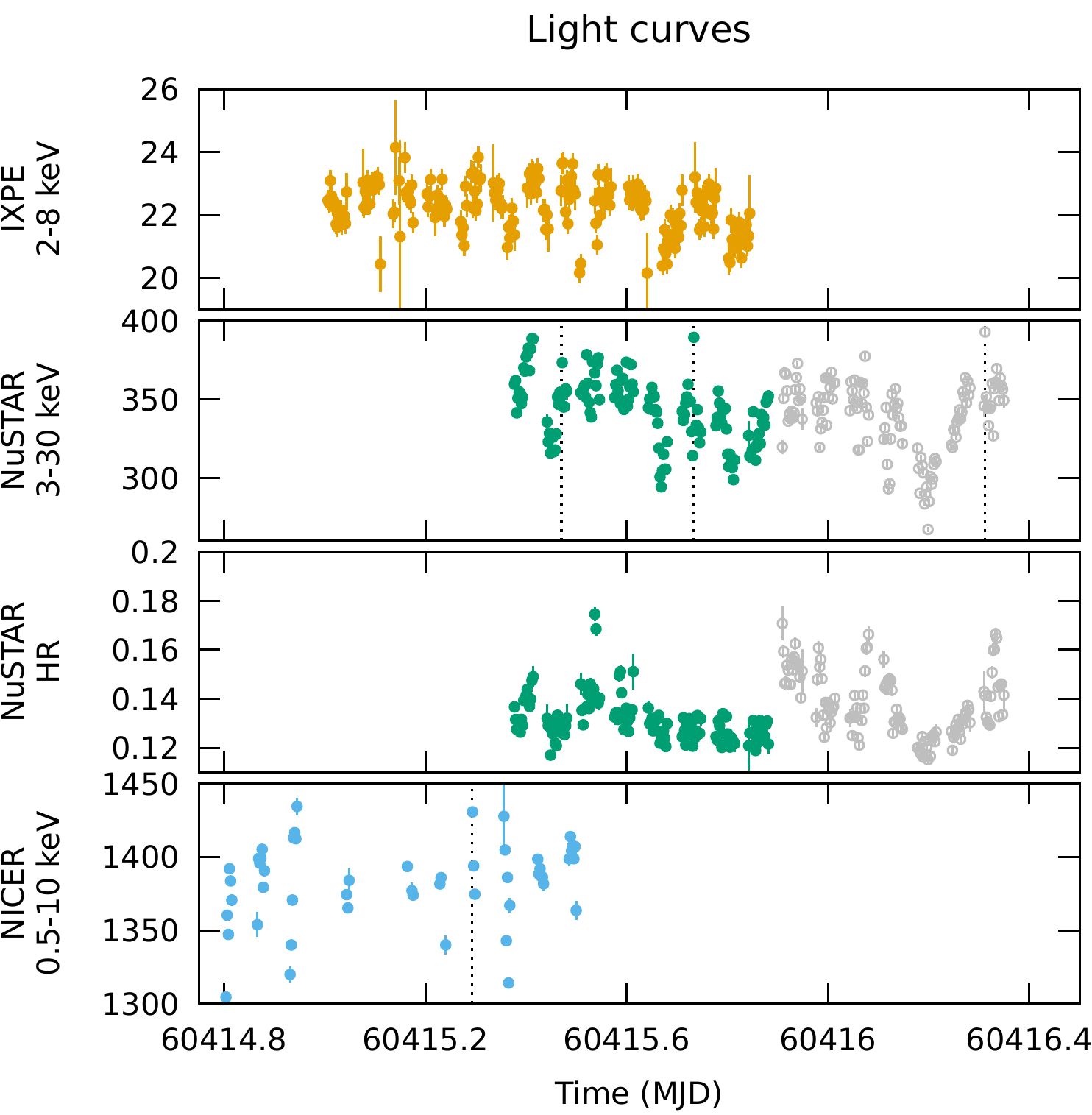}
\caption{
\ixpe, \textit{\nustar}, and \nicer\ light curves of Ser~X-1 (count\pers). The third panel shows the \textit{\nustar}\ hardness ratio, (10--30 keV)/(3--10 keV).
 Time bins of 200 s are used. Empty gray circles denote the \textit{\nustar}\ points not simultaneous with the \ixpe\ exposure. Dotted lines mark the type-I bursts. 
}
\label{fig:lc}
\end{figure}

\begin{figure*}
\includegraphics[height=5cm]{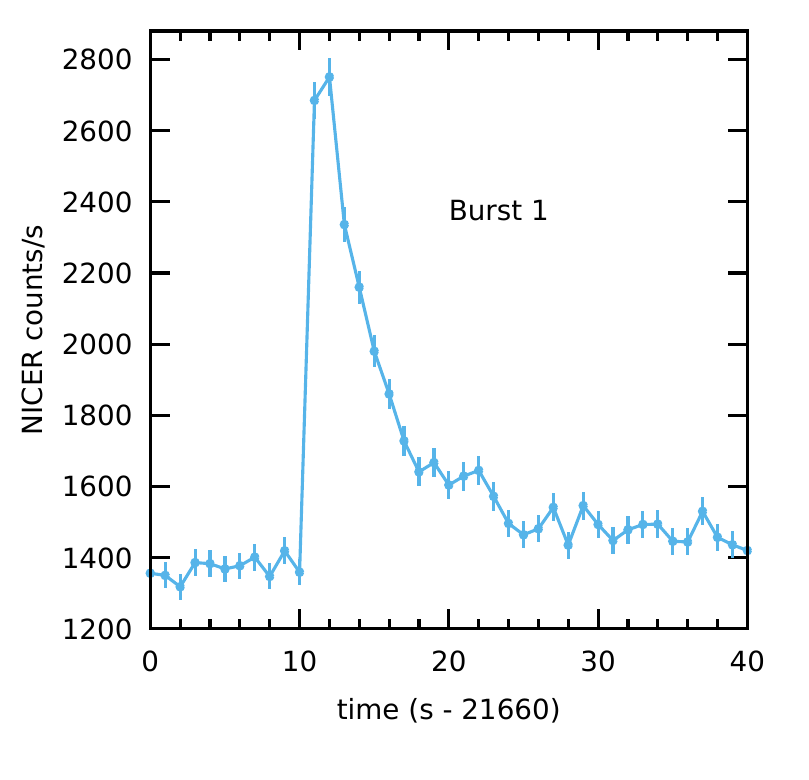}
\includegraphics[height=5cm]{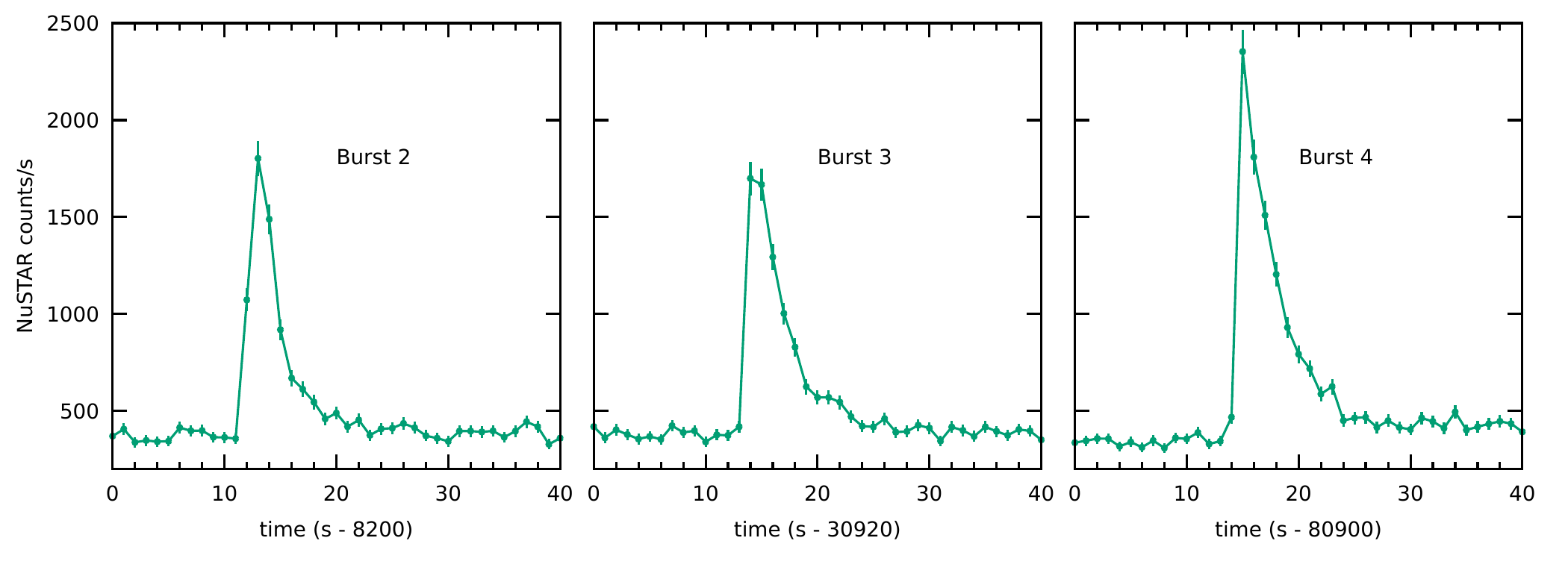}
\caption{
Profiles of the four type-I X-ray bursts detected with \nicer\ (left panel) and \textit{\nustar} (right panels). The first two occurred during the \ixpe\ exposure.
}
\label{fig:bursts}
\end{figure*}

The 200-second-binned light curves obtained with \ixpe, \nicer,\, and \textit{\nustar}\ are shown in Fig.~\ref{fig:lc}.
The flux variability is apparent, albeit not dramatic. However, some spectral variability is also observed. In particular, the \textit{\nustar}\ hardness, 
(10--30 keV)/(3--10 keV),
changes more rapidly during the second half of the \textit{\nustar}\ exposure, right after the end of the \ixpe\ exposure. Thus, for the sake of the joint spectropolarimetric fit, we only used the \nicer\ and \textit{\nustar}\ data that were taken simultaneously with the \ixpe\ exposure. 

We also constructed 1-second-binned light curves to search for X-ray bursts. We found clear evidence for four type-I X-ray bursts, one observed with \nicer\  and the other three with \textit{\nustar}. These four bursts are plotted in Fig.~\ref{fig:bursts}. The recurrence times of the bursts are irregular, between 4.5 and 13 hours, consistent with past observations of the source \citep{sztajno1983,galloway2008}. Only the first and second bursts occurred during the \ixpe\ exposure. The third burst occurred during a gap between \ixpe\ observation segments, and the fourth burst occurred after the end of the \ixpe\ exposure. All bursts are well described by a fast rise and exponential decay profile, with an e-folding decay time of 3--4~s \citep[see also][]{galloway2008}.

We investigated the differences between burst and non-burst emission as follows.
We extracted the burst and non-burst data for each instrument; since the \ixpe\ data alone do not provide a sufficient signal-to-noise to identify the bursts, we defined the \ixpe\ good time intervals (GTIs) from the \nicer\ and \textit{\nustar}\ light curves. In other words, for \ixpe\ we extracted the data of the first and second bursts using the \nicer\ and \textit{\nustar}\ GTIs. 
For each burst, the GTIs were created in 10-second intervals starting from the burst onsets, as determined from the light curves (Fig.~\ref{fig:bursts}).
For the polarimetric analysis, we extracted a single burst spectrum, combining the two intervals. 
We note that these two bursts have similar profile and spectral parameters (see also Appendix \ref{sec:Bursts}).
In the following step, we analyzed the Stokes parameters for the whole \ixpe\ exposure, the non-burst period, and the bursts.

\subsection{Polarimetric analysis}

\begin{figure}
\includegraphics[width=1.0\columnwidth]{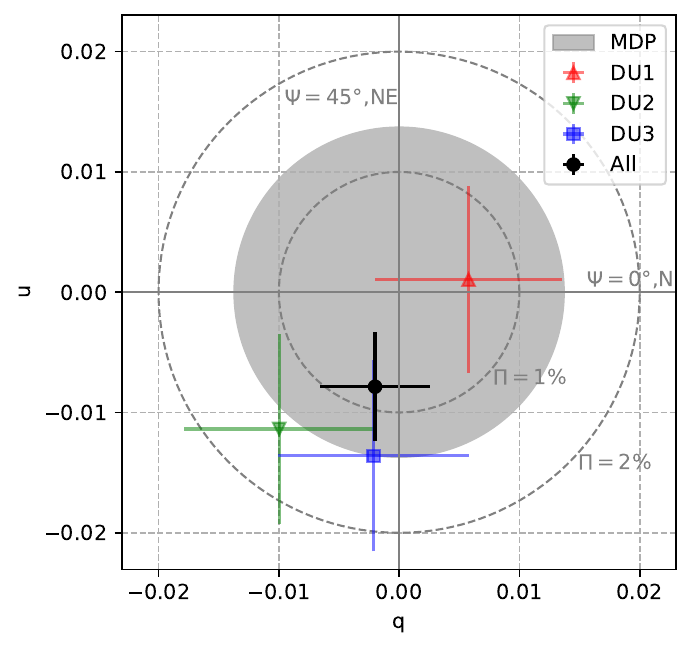}
\caption{
Normalized Stokes $q$ and $u$ parameters in the 2--8 keV band for the three \ixpe\ DUs and their combination. The gray-filled circle encloses the 99\% MDP.
}
\label{fig:Stokes}
\end{figure}

\begin{figure}
\includegraphics[width=1.0\columnwidth]{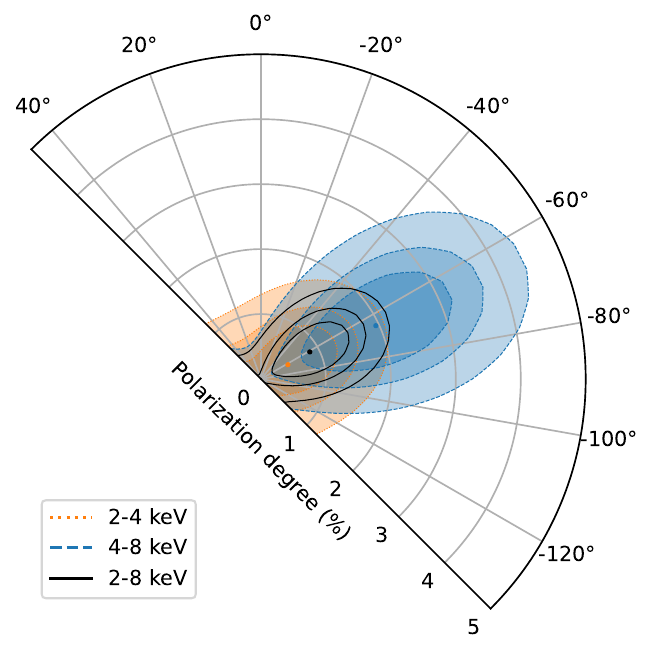}
\caption{
Contour plots of the PD and PA at the 68, 90, and 99\% confidence levels, in the 2--4 keV (dotted orange), 4--8 keV (dashed blue), and 2--8 keV (solid black) energy bands. 
Color transparency increases with increasing confidence level. 
}
\label{fig:contours}
\end{figure}

We first performed a polarimetric analysis of the whole \ixpe\ observation (i.e., not separating burst and non-burst periods).
The normalized Stokes parameters measured by \ixpe\ in the source region are plotted in Fig.~\ref{fig:Stokes}, together with the minimum detectable polarization (MDP) at the 99\% level. We do not obtain a significant detection, with an upper limit to the 2--8 keV polarization degree (PD) of 2\% at the 99\% confidence level (one degree of freedom)\footnote{
The quoted upper limits on the PD are derived from the one-dimensional error, i.e., independent of the value of the PA. 
}. We do not observe any significant evolution with energy. 
In Table~\ref{tab:poldeg} we report the PD measured in different energy bands with \xspec\ 12.14.0 \citep{arnaud1996}.
In Fig.~\ref{fig:contours} we plot the two-dimensional contours of the  PD and polarization angle (PA) measured with \xspec. 
The statistical significance is around 68\% in the whole 2--8 keV band and increases up to 90\% when considering just the  4--8 keV band, however not enough to warrant a detection claim.
    We obtain consistent values from the polarization cubes extracted with \textsc{ixpeobssim} \citep{baldini2022}.  

We also performed a similar analysis of the \ixpe\ data of the burst phases. Since the exposure amounts to just 20 s, the MDP is 80\%, and the upper limit to the burst polarization is on the same order. 

\begin{table}
\caption{\label{tab:poldeg}
PD and PA measured with \textsc{xspec}.
}
\begin{center}
  \begin{tabular}{cccc}
    \hline    \hline
 Energy range (keV)   & PD (\%)  & PD U.L. (\%) & PA (\degr) \\
 \hline 
2--8 &  $0.8 \pm 0.4$ & $<2.0$ & $-60 \pm 15$ \\
2--4  &$0.5 \pm 0.5$ & $<1.9$ & - \\
4--8  &$1.9 \pm 0.8$ & $<4.0$ & $-64 \pm 12$ \\
\hline 
\end{tabular}
\tablefoot{
Uncertainties are given at the 68\% ($1 \sigma$) confidence level for one parameter of interest; upper limits are quoted at the 99\% confidence level.
}
\end{center}
\end{table}

\subsection{Spectropolarimetric analysis}

We performed a joint fit of \ixpe, \nicer,\ and \textit{\nustar}\ spectropolarimetric data with \xspec. 
Here we focus on the stationary (non-burst) phase; the spectra of the bursts are discussed in Appendix~\ref{sec:Bursts}.
To take interstellar absorption into account, we always included the \tbabs\ model \citep[][]{tbabs}, leaving the column density as a free parameter. We included cross-calibration constants, and we also applied a gain correction to the \ixpe\ and \nicer\ spectra using the \texttt{gain fit} command in \xspec\ to account for the imperfect spectral agreement between the different instruments.
We also found some residuals in the \nicer\ spectrum below 2 keV, likely due to calibration issues\footnote{\url{https://heasarc.gsfc.nasa.gov/docs/nicer/analysis\_threads/arf-rmf/}} \cite[see also][]{fabiani2024}, and we included an absorption edge to account for them. 

\begin{figure*}
\includegraphics[width=\columnwidth]{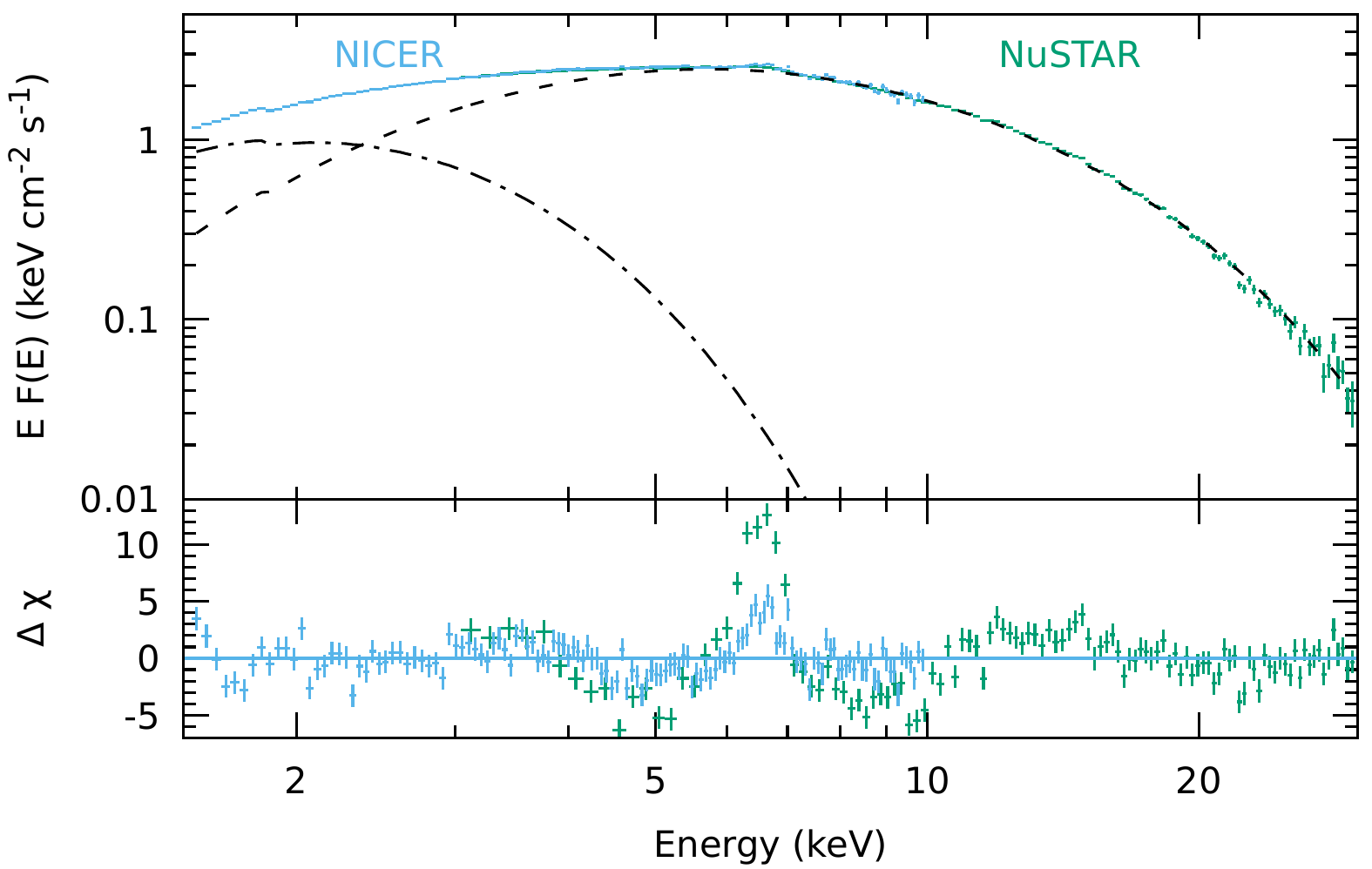}
\includegraphics[width=\columnwidth]{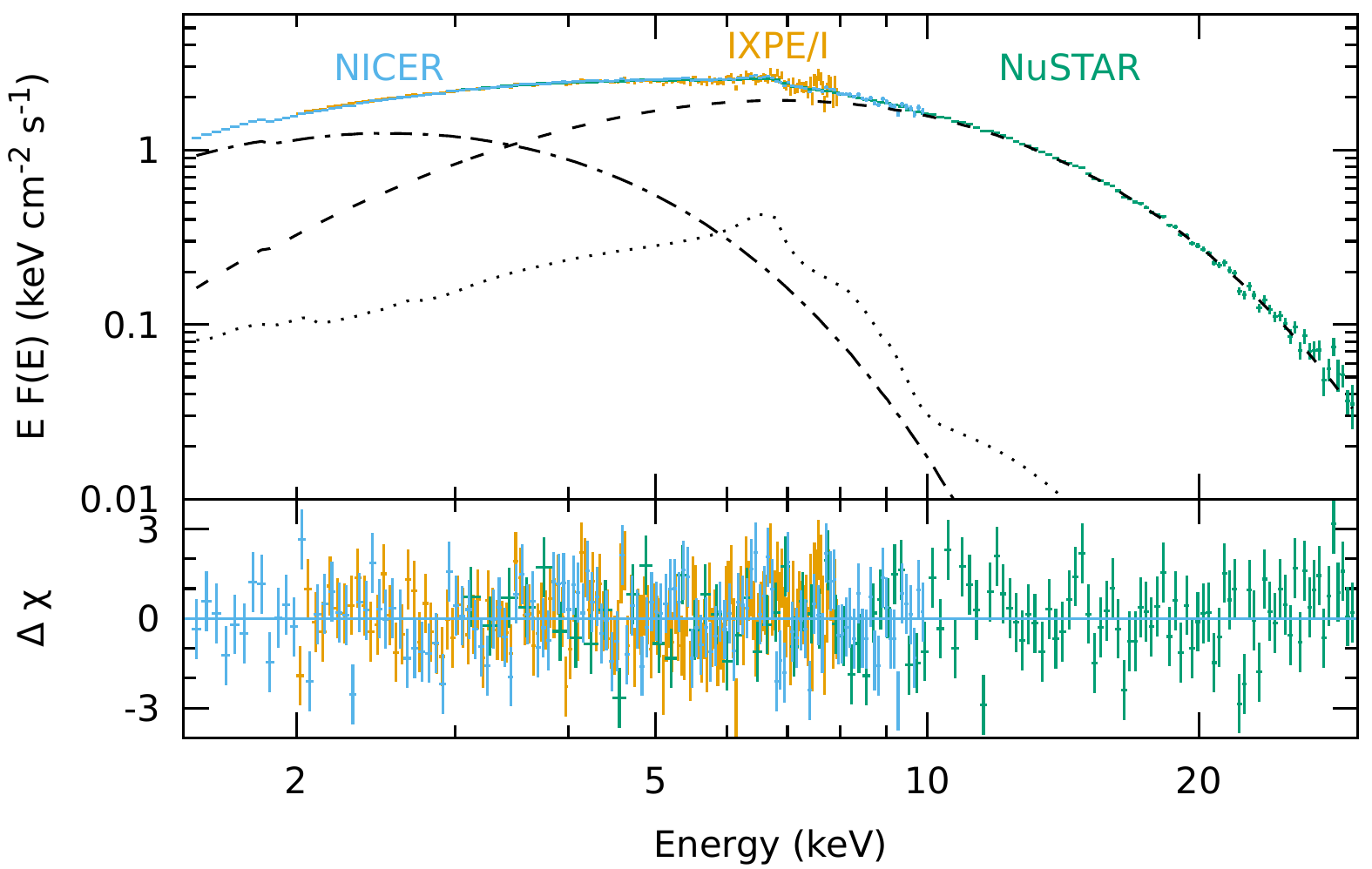}
\caption{
Deconvolved flux spectra fitted with different models and corresponding residuals.
Left panel: Model composed of \diskbb\ (dash-dotted line) and \thcomp*\bbodyrad\ (dashed line), with residuals showing the presence of reflection features. For the sake of clarity, only \nicer\ and \textit{\nustar}\ data are shown. 
Right panel: Model including \relxillns\ (dotted line). The data from \textit{\nustar}/FPMA and FPMB were combined (with \texttt{setplot group} in \xspec) for plotting purposes only.
}
\label{fig:eeuf_del}
\end{figure*}

First, we fitted the \ixpe, \nicer,\ and \textit{\nustar}\ spectra with a model consisting of two components: a disk multicolor blackbody \citep[\diskbb\ in \xspec;][]{mitsuda1984} and a Comptonized blackbody (\bbodyrad\ convolved with \thcomp; \citealt{thcomp}). The model was multiplied by a constant polarization via \polconst. This fit is, however, unacceptable, as the \nicer\ and \textit{\nustar}\ residuals clearly show the presence of a strong iron line and Compton reflection (see Fig. \ref{fig:eeuf_del}, left panel). 

Then, we included a reflection component with \relxillns\ \citep{relxillns}. The final model thus reads 
$$
\polconst(\diskbb+\thcomp*\bbodyrad+\relxillns).
$$
In \relxillns\ we fixed the dimensionless spin $a=0.2$
\citep[assuming a typical NS spin frequency; see][]{braje2000}
and the number density $\log n_{\rm e}=18$ \citep[appropriate for the inner disk of a NS-LMXB; see, e.g.,][]{relxillD} because these parameters are not well constrained by the fit. 
With the addition of the reflection component, the fit eventually achieves an acceptable value with $\chi^2$/d.o.f. $= 1031/962$ (see Fig. \ref{fig:eeuf_del}). The extrapolated luminosity in the 0.1--100 keV band is $7.6 \times 10^{37}$ \lumcgs, corresponding to an Eddington ratio of $0.28(1+X),$ where $X$ is the hydrogen mass fraction. 
The best-fitting parameters are reported in Table \ref{tab:Fit}.

\begin{table}
\caption{\label{tab:Fit}
Best-fitting model parameters of the fits to the \textit{\nustar}\ and \textit{\nustar}+\ixpe\ data. }
\begin{center}
  \begin{tabular}{ll}
\hline
\hline
Parameter & \\ 
\hline 
   \multicolumn{2}{c}{\diskbb} \\
$kT_{\rm in}$ (keV) & $0.93^{+0.01}_{-0.03}$ \\
$R_{\rm in} \sqrt{\cos i}$ (km) & $13.1^{+0.2}_{-0.4}$   \\
\hline
 \multicolumn{2}{c}{\bbodyrad} \\
 $kT_{\rm bb}$ (keV) & $1.38^{+0.02}_{-0.05}$ \\
 $R_{\rm bb}$ (km) & $7.7^{+0.2}_{-0.5}$ \\
 \hline
 \multicolumn{2}{c}{\thcomp} \\
 $\tau$ & $8.1^{+0.3}_{-0.2}$ \\
 $kT_{\rm e}$ (keV) & $2.92^{+0.03}_{-0.04}$ \\
 \hline
\multicolumn{2}{c}{\relxillns} \\
$q_{\rm em}$ & $2.26 \pm 0.09$ \\
$a$ & [0.2] \\
incl (deg) & $25.1^{+1.3}_{-1.5}$ \\
$R_{\rm in}$ (units of $R_{\rm ISCO}$)  & $1.6^{+0.3}_{-0.5}$ \\
$kT_{\rm bb}$ (keV) & [=$kT_{\rm bb, \bbodyrad}$] \\
$\log (\xi$/erg cm \pers) & $2.88\pm 0.07$ \\
$A_{\rm Fe}$ & $5.0^{+1.1}_{-0.3}$\\
$\log n_{\rm e}$ & [18] \\
 $N_{\rm r}$ ($10^{-3}$) & $1.37^{+0.09}_{-0.08}$ \\
 \hline
\multicolumn{2}{c}{Cross-calibration} \\
$C_{\rm DU1-FPMA}$ & $0.881^{+0.005}_{-0.007}$ \\
$C_{\rm DU2-FPMA}$ & $0.880^{+0.006}_{-0.005}$ \\
$C_{\rm DU3-FPMA}$ & $0.861^{+0.005}_{-0.006}$ \\
$C_{\rm FPMB-FPMA}$ & $0.987 \pm 0.001$ \\
$C_{\rm XTI-FPMA}$ & $0.929 \pm 0.001$ \\
\multicolumn{2}{c}{Gain shift} \\
$\alpha_{\rm DU1}$ & $1.013^{+0.002}_{-0.003}$ \\
$\beta_{\rm DU1}$ (eV) & $-53^{+12}_{-10}$ \\
$\alpha_{\rm DU2}$ & $1.008^{+0.002}_{-0.003}$ \\
$\beta_{\rm DU2}$ (eV) & $-30^{+12}_{-11}$ \\
$\alpha_{\rm DU3}$ & $1.007^{+0.002}_{-0.003}$ \\
$\beta_{\rm DU3}$ (eV) & $-33^{+12}_{-11}$ \\
$\alpha_{\rm XTI}$ & $1.032\pm 0.001$ \\
$\beta_{\rm XTI}$ (eV) & $-65^{+2}_{-3}$ \\
\hline
$\chi^2$/d.o.f. & 1031/962\\
\hline
\multicolumn{2}{c}{Photon flux ratios\tablefootmark{a}} \\
\multicolumn{2}{c}{2--8 keV} \\
$F_{\rm diskbb}/F_{\rm tot}$ & 0.44\\
$F_{\rm bbodyrad}/F_{\rm tot}$ & 0.47 \\
$F_{\rm relxillns}/F_{\rm tot}$ & 0.09 \\
\multicolumn{2}{c}{2--4 keV} \\
$F_{\rm diskbb}/F_{\rm tot}$ & 0.58 \\
$F_{\rm bbodyrad}/F_{\rm tot}$ & 0.35 \\
$F_{\rm relxillns}/F_{\rm tot}$ & 0.07 \\
\multicolumn{2}{c}{4--8 keV} \\
$F_{\rm diskbb}/F_{\rm tot}$ & 0.19 \\
$F_{\rm bbodyrad}/F_{\rm tot}$ & 0.69
\\
$F_{\rm relxillns}/F_{\rm tot}$ & 0.12 \\
\multicolumn{2}{c}{Energy flux (2--8 keV)} \\
$F_{\rm tot}$ (\fluxcgs) & $5.1\times 10^{-9}$ \\
\hline 
\end{tabular}
\tablefoot{
Uncertainties are given at the 68\% confidence level for one parameter of interest. Parameters in square brackets were kept frozen during the fit. The normalizations of \diskbb\ and \bbodyrad\ are computed assuming a source distance of 7.7 kpc.\\
\tablefoottext{a}{The photon fluxes are in units of photon\percm\pers.}
}
\end{center}
\end{table}

Finally, we conducted some tests making different assumptions regarding the polarization of the various spectral components. We fitted the data using a similar model as above, with the only difference being that each component is multiplied by a different \polconst. The results are reported in Table~\ref{tab:pol_tests}.
In these tests, we assumed that the PA of the Comptonized blackbody is perpendicular to that of the disk emission. This choice is consistent with the results obtained for Cyg~X-2 \citep{farinelli2023} and with the hypothesis that the BL or SL is vertically extended. We also assumed that the PA of the reflection component is the same as the Comptonized blackbody.
We first fixed the PD of both the disk and Comptonized component to 1\%, which is consistent with the low inclination of the source; in this case, we obtain an upper limit to the PD of the reflection component of 22\%. Then, we fixed the reflection PD to 10\% and left only one of the other two components free to vary, obtaining upper limits of 3.4\% to 3.6\%. Next, we assumed a zero PD for the Comptonized component and found an upper limit of 1.7\% for the disk PD. Finally, we tested a scenario consistent with the numerical simulations by \cite{lapidus&sunyaev1985}, namely, assuming a PD of 3\% for the Comptonized plus reflected emission; in this case, we find an upper limit of 3.8\% for the disk PD. 
We remark that all these tests provide the same goodness of fit.

\begin{table}
\caption{\label{tab:pol_tests}
PD and PA of each spectral component for different scenarios. 
}
\begin{center}
  \begin{tabular}{lll}
\hline
\hline
Component & PD (\%) & PA (deg) \\ 
\hline 
\diskbb & [1] & =PA$_{\bbodyrad} - 90$\\
\thcomp*\bbodyrad & [1] & (unconstrained) \\
\relxillns & $<22$ & =PA$_{\bbodyrad}$ \\
\hline
\diskbb & [1] & =PA$_{\bbodyrad} - 90$\\
\thcomp*\bbodyrad & $<3.4$ & (unconstrained) \\
\relxillns & [10] & =PA$_{\bbodyrad}$ \\
\hline
\diskbb & $<3.6$ & =PA$_{\bbodyrad} - 90$\\
\thcomp*\bbodyrad & [1] & (unconstrained) \\
\relxillns & [10] & =PA$_{\bbodyrad}$ \\
\hline
\diskbb & $<1.7$ & =PA$_{\bbodyrad} - 90$\\
\thcomp*\bbodyrad & [0] & - \\
\relxillns & [10] & (unconstrained) \\
\hline
\diskbb & $<3.8$ & =PA$_{\bbodyrad} - 90$\\
\thcomp*\bbodyrad & [3] & (unconstrained) \\
\relxillns & [3] & (unconstrained) \\
\hline
\end{tabular}
\tablefoot{Upper limits are reported at 99\% confidence level for one interesting parameter ($\Delta \chi^2=6.63$). Parameters in square bracket are frozen. 
}
\end{center}
\end{table}

\section{Discussion and conclusions}\label{sec:conclusions}
The joint \ixpe/\nicer/\textit{\nustar}\ observation of Ser~X-1 -- and especially its comparison with other NS-LMXBs observed by \ixpe\ so far --
 allows us to draw interesting conclusions.
Our analysis confirms the presence of a relativistic reflection component; this gives an inclination angle of 25\degr, which is less extreme than the values of $<$10\degr\ reported by \citet{miller2013} and \citet{ludlam2018_SerX1} and consistent with the values reported by \citet{cackett2010}, \citet{chiang2016_a,chiang2016_b}, \citet{matranga2017}, and \citet{mondal2020_SerX1}. The iron line requires an iron abundance of 5 times the solar one, consistent with the value obtained by \citet{ludlam2018_SerX1}.
From the reflection component, we also derive a disk truncation radius of the innermost stable circular orbit (ISCO) between 1.1 and 1.9, or 14--24 km assuming a canonical NS mass of 1.4~$M_{\odot}$. This is not far from the radius of the disk blackbody component, which is found to be in the range 12--13 km for an inclination of 25\degr.
Polarized reflected emission has been suggested as a possible explanation for the unexpectedly large polarization observed in some atolls \citep{saade2024,dimarco2023_4U1820} as well as Z sources, especially when observed in the horizontal branch \citep{cocchi2023,fabiani2024}. However, for Ser~X-1  we do not obtain a detection despite the significant reflection component.

The two atolls GS~1826$-$238 and GX~9+9 provide an interesting comparison with Ser~X-1. GS~1826$-$238 has a low polarization ($<1.3\%$) and no reflection features \citep{capitanio2023}. GX~9+9, on the contrary, shows a significant reflection component \citep{iaria2020} and a PD of $\sim$2\%  \citep{chatterjee2023,ursini2023}, which can be attributed to a combination of BL and reflected emission \citep{ursini2023}. 
Ser~X-1 is a somewhat mixed case, in that the reflected emission is significant but the X-ray polarization is relatively low ($<2\%$). We can attribute this to the low inclination of the source. Indeed, numerical simulations indicate that the polarization of the radiation emitted by the BL is $\lesssim 1\%$ \citep{gnarini2022, farinelli2024}. A low polarization is also expected for the intrinsic disk emission at low inclinations \citep{chandra1960}. The only other component that could produce a significant polarization is Compton reflection \citep{matt1993,poutanen1996,schnittman&krolik2009}. \citet{lapidus&sunyaev1985} calculated the polarized fraction of X-ray-burster radiation, both during bursts and in the stationary phase, for the BL plus disk reflection. 
The authors found a relatively large polarization in stationary phases, even for low inclinations ($\sim 3\%$ at 25\degr). However, their calculations did not include the intrinsic disk emission, which in Ser~X-1 accounts for 44\% of the 2--8 keV photon flux (see Table \ref{tab:Fit}) and can dilute the polarization. It is indeed possible to fit the data assuming a 3\% PD of the BL plus reflection component, in which case the fit yields an intrinsic disk polarization lower than 3.8\%.

We note that, at the accretion rate of Ser~X-1 ($\sim 30\%$ of the Eddington limit; see Sect. 3.3), the BL is indeed expected to form a bright equatorial belt \citep{inogamov&sunyaev1999,popham&sunyaev2001,suleimanov&poutanen2006}. It is reasonable to assume that a fraction of the BL emission is reflected by the disk, whose inner radius is not too far from the NS surface, and becomes polarized. However, different configurations are possible. For example, assuming the extreme case of a zero polarization of the high-energy component, we get an upper limit to the disk polarization of 1.7\%, which would still indicate an inclination lower than 50°. 

Finally, four type-I short X-ray bursts were detected during the observation. Their properties (occurrence, profile, and duration) are consistent with previous observations of Ser~X-1 with \rxte\ \citep{galloway2008}. 
The upper limit to the burst polarization is 80\%.
It might be possible to obtain better constraints on the burst polarization by stacking frequent long bursts, which tend to occur at low accretion rates (i.e., in the low hard state). 
Future \ixpe\ observations of NS-LMXBs, possibly in different spectral states, will place further constraints on their average polarization properties and extend our knowledge of these sources. 

\begin{acknowledgements}

This research used data products provided by the IXPE Team (MSFC, SSDC, INAF, and INFN) and distributed with additional software tools by the High-Energy Astrophysics Science Archive Research Center (HEASARC), at NASA Goddard Space Flight Center (GSFC). The Imaging X-ray Polarimetry Explorer (IXPE) is a joint US and Italian mission.  
FU, AG, SB, FC, and GM acknowledge financial support by the Italian Space Agency (Agenzia Spaziale Italiana, ASI) through contract ASI-INAF-2022-19-HH.0.
SF has been supported by the project PRIN 2022 - 2022LWPEXW - “An X-ray view of compact objects in polarized light”, CUP C53D23001180006.
AB acknowledges support from the Finnish Cultural Foundation grant 00240328. 
JP thanks the Academy of Finland grant 333112 for support. 
\end{acknowledgements}

\bibliographystyle{aa} 
\bibliography{biblio} 

\begin{appendix}
\section{Spectral analysis of the bursts}\label{sec:Bursts}
To perform a time-resolved spectral analysis of the bursts, we extracted the spectra within intervals of 1 s, covering the duration of the bursts. We fitted the data subtracting the persistent emission as background. This standard procedure is based on the assumption that the persistent emission does not change in the burst phase, which is often a good approximation (e.g., \citealt{vanparadijs&lewin1986}, \citealt{galloway2008}). We modeled the burst emission with a simple blackbody. 
We have also attempted more sophisticated treatments, such as scaling the background with a free constant \citep[e.g.,][]{worpel2013,worpel2015,sanchez2020} or including a further reflection component \citep[e.g.,][]{bult2019}. However, the fits are not improved and the free parameters become poorly constrained.

The results are summarized in Table \ref{tab:bursts}. The peak temperature of each burst is around 2 keV. The peak blackbody radius is 8--9 km assuming a distance of 7.7 kpc, and we do not find evidence for photospheric radius expansion. The average peak flux is $(20 \pm 2) \times 10^{-9}$\fluxcgs, consistent with the value reported for this source by \cite{galloway2008}. 

\begin{table}
\caption{\label{tab:bursts}
Spectral parameters at the peak for each burst. 
}
\begin{center}
  \begin{tabular}{lcccc}
\hline
\hline
Burst & $kT_{\rm bb,peak}$ & $R_{\rm bb, peak}$ & $F_{\rm bb,peak}$ & $\chi^2$/d.o.f. \\
&(keV)&(km)&($10^{-9}$ \fluxcgs) \\
\hline
1 & $1.92^{+0.13}_{-0.12}$ &$8.2 \pm 0.6$&$16 \pm 3$ & $89/73$ \\
2 & $1.95^{+0.16}_{-0.14}$ & $8.4^{+1.3}_{-1.1}$ & $18.4 \pm 1.5$ & $53/46$ \\
3 & $2.10^{+0.19}_{-0.16}$ & $7.3 \pm 1.1$ & $19.7 \pm 1.7$ & $48/42$ \\
4 & $2.00^{+0.14}_{-0.12}$ & $9.7^{+1.2}_{-1.1}$ & $27.3 \pm 1.8$ & $52/52$\\
\hline
\end{tabular}
\end{center}
\end{table}

\end{appendix}

\end{document}